# Mechanical Design, Control Choices and first Return of Use of a Prosthetic Arm

G. Thomann[*]  
Sols, Solids, Structures Laboratory  
INP Grenoble, France

V. Artigues[†]  
Tech.Innovation  
Evry, France

**Abstract**—In the world of upper limb prostheses, few companies dominate the majority of the market. They propose different kinds of hand, wrist and elbow prostheses but their control is often difficult to understand by the patients.
We have decided[*] to develop new myoelectric prosthetic arm (elbow, wrist and hand) by axing our development on the use of new technologies and facility of use for the patient.
In this paper, we are explaining in details the different kinds of prostheses currently proposed to the amputees, their advantages and their drawbacks, the descriptions of the patients' needs and the possible improvements of the product. We will develop the designing choices of our prosthesis and the movements it can realize. Then we will explain the simplified control of the product by the patient and its first reactions. Finally, we will conclude by the news ideas and the next researches to concretize.

## I. Introduction

The aim of the product developed is to improve the conditions of life of the amputees and to help them to find (again) independence and dignity. The prostheses are designing to be used by upper limb amputees – elbow, wrist and hand.

Unfortunately, current high tech prostheses are too expensive and it is difficult for a patient to buy an adapted prosthesis (to its needs and desires). By using new technologies, our aim is to propose a new tool for a lower cost and a better comfort of use to the amputees.

Thanks to a first study, to be used and appreciate by the patient, the prosthesis must be functional, aesthetics, quiet, easy to use, light and understood.

The aim of the developed product is to help amputees and to propose them low cost solutions for a better use. For these reasons, the state of the art is focused on the upper limb prosthesis market.

Before introducing our proposal prostheses, we are going to detail the different prosthesis developed by the main societies. UTAH, OTTA BOCK and PROTEOR Societies propose three products on the market. These three companies concentrate 90% of the market.

*E-mail: guillaume.thomann@hmg.inpg.fr  
† E-mail: vincent.artigue@techinnovation.fr

Currently, three kinds of prostheses are proposed on the market, we will detail them in the following section:
- aesthetics prosthesis,
- mechanical prosthesis,
- myoelectric prosthesis.

### A. Aesthetics Prostheses

Their aim is only aesthetics, and this type of prosthesis is generally used by patients. In the majority of the cases, the prosthetic arm is created from a standard mould. It means that the resemblance to the healthy member is not optimal. This type of prosthesis does not carry out any movement; it is applied to the patient simply to restore its body. The pictures below show an aesthetics hand from OTTO BOCK [1] (figure 1) and a complete aesthetics prosthesis (figure 2).

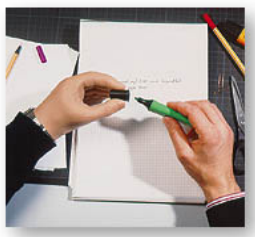
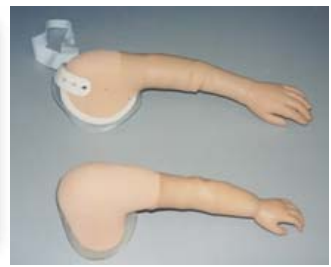

Fig. 1. An aesthetics hand from OTTO BOCK Society
Fig. 2. Picture of complete aesthetics prosthesis

### B. Mechanical Prostheses

The mechanical prostheses try to approach the functionality of the old member. They can be manual (use with the assistance of the healthy member) or with cable with the use of a harness.

Three kinds of mechanical elbow products are currently offered to the patients.

The first is the elbow with toothed rack (figure 3), which is released thanks to a pushbutton actuated by the valid hand or by a cable.

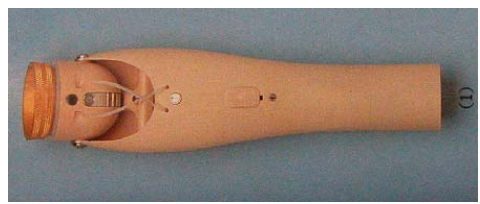

Fig. 3. OTTO BOCK elbow with toothed rack

The front arm and the elbow are provided; three sizes of this product are proposed by this company. There are many drawbacks: noise of the toothed rack, the limited number of positions of the front arm and the bad aesthetic of the pushbutton.

The second elbow is the elbow with friction, which moves thanks to the friction of a spiral spring on the axis of the elbow [2]. A cable ordered by the other shoulder actuates blocking: traction locks it, another unbolts it. It is more functional than the precedent, but maintains the position less firmly. In addition, it needs a double order from the amputee, which is not always easy to carry out.

Lastly, there is an automatic elbow from OTTO BOCK (figure 4). The front arm is manufactured out of plastic and is not very solid. Its distal part (near to the wrist) is cylindrical and is simply cut to the length of the healthy member. Unfortunately, prosthetic arm will not resemble to the healthy member.

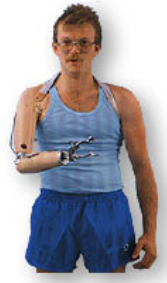

Fig. 4. Mechanical elbow from OTTO BOCK

Being informed of all the drawbacks of these mechanical prostheses and having the technologies to improve them, our objectives are to propose to the patients a more functional mechanical prosthesis and solutions compared to the criteria of aesthetics, quiet, easy to use, light and understood.

*C. Myoelectric Prostheses*

Myoelectric signals (Electromyogram or EMG) are electrical signals registered from the muscles activities. Thanks to these signals, a great number of applications are possible. The EMG are complex signals with noise and easily influenced by many factors. Then, from the interpretation to the use, the EMG need several specific treatments [3].

With surface electrodes placed directly on the skin, it is easily possible to measure functional motor activities such as washing the teeth or writing [4].

Now, it becomes possible to control computers without joysticks or keyboards. An experiment to demonstrate bioelectric flight control of 757 class simulation aircraft landing at San Francisco International Airport has been tested [5]. A pilot closes a fist in empty air and performs control movements which are captured by a dry electrode array on the arm, analyzed and routed through a flight director permitting full pilot outer loop control of the simulation.

The consequences of the Viet Nam war were in the origin of the development of the UTAH products. This society was the first to propose the EMG technology to control the prosthesis. The picture below (figure 5) shows EMG prosthesis of elbow from the UTAH Company.

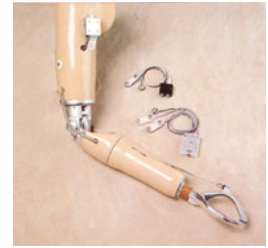

Fig. 5. EMG arm prosthesis proposed by UTAH

The OTTO BOCK society proposes prosthesis of hand coupled with a mechanical elbow (figure 6) [6]. The hand is a tree legs grip with an aesthetic glove.

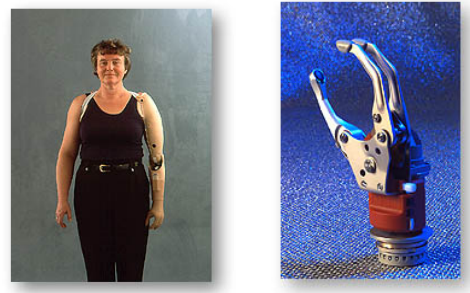

Fig. 6. Arm prosthesis proposed by OTTO BOCK

*D. Conclusion*

This state of the art of the prostheses shows that there are many problems. The most important is the average cost of the myoelectric prosthesis: 23 k€ for the elbow, 7.5 k€ for the hand and its wrist. The high prices of these prostheses explain the weak diffusion and use.

A great number of myoelectric prostheses are criticized by amputees because of the less of aestheticisms and the difficulty of control: co-contraction to control the prostheses is too difficult). Some of them are noisy (mechanical prosthesis) and sometimes the prosthesis doesn't answer as the patient wants.

For all these reasons, we want to improve the functionality of the myoelectric prostheses as their aesthetic aspect, by using last technologies as regards machining, electronics and mechanics.

The project of developing myoelectric upper limb prosthesis was born in 1998. Accordingly, Tech.Innovation society was created in 2000. Its aim is to be the first to propose functional and aesthetic myoelectric prostheses, completely refunded by the sickness insurances.

## II. Design of the Prosthesis

In this part, we are going to explain the economical design choices of the prosthesis to improve the quality, the aestheticism, the functionality and the weight.

The advantage of this new prosthesis comes from its new central geometrical structure (figures 7a and 7b), which integrates much of the elements: the motors, the batteries, the bevel gearbox, the weight compensator, the control card and the hand.

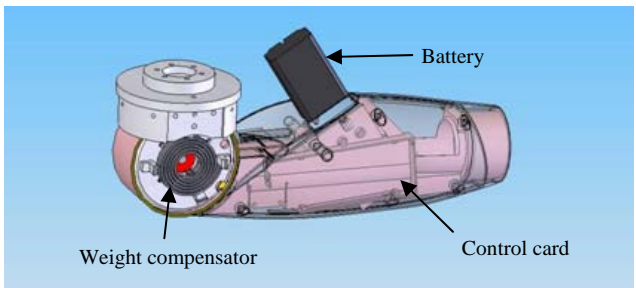

Fig. 7a. First view of the central structure of the prosthesis

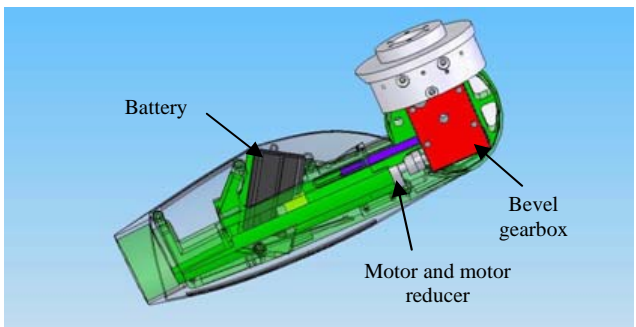

Fig. 7b. Second view of the central structure of the prosthesis

Moreover, with these characteristics, it is possible to test the different functions of the prosthesis without positioning the aesthetic masks.

Thanks to a simple mechanism, the patient can manually order the positioning of the elbow. Moreover one button was integrated in order to manually open the hand if the patient does not manage to order his prosthesis.

A maximum of the parts of this prosthesis are designing in CAD and build thanks to stereolithiography process, which an economical process. The advantages of this process are the lightness and the solidity but also the possibility to build the shape wanted. It is then possible to propose a prosthesis shape nearest to the healthy member shape. The intern structure of the prostheses (shown in figure 7) is the same.

After reconstruction, the result of the prosthesis without the wrist and the hand is shown in figure 8.

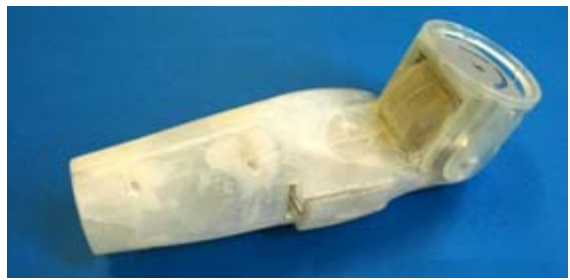

Fig. 8. Picture of the body part of the prosthesis with the elbow: use of Stereolythography process

The OTTO BOCK society uses its own battery; the consequence is an increasing price of de prosthesis.

We have decided to choose a Lithium battery (the same that commercial ones). This choice enabled us to divide by two the weight of the batteries and to reduce the production costs.

Concerning the hand, the gripping of the objects with a standard three legs grip hand (figure 9) is used with an aesthetics glove. For the moment, the catch of object is difficult and its position is unstable but the future version will integrate passive compliance.

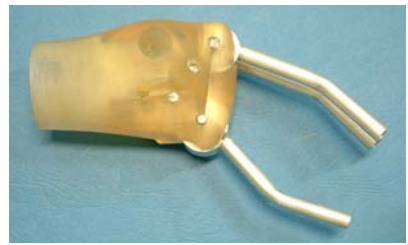

Fig. 9. The standard three legs grip hand used

Finally, the figure 10 shows the current prosthesis proposed by Tech.Innovation Society. In this figure, we can easily locate the fit, the elbow, the wrist, the hand and the body part (operating part). The fit allows the adaptation between the patient and the prosthesis; it integrates the two electrodes.

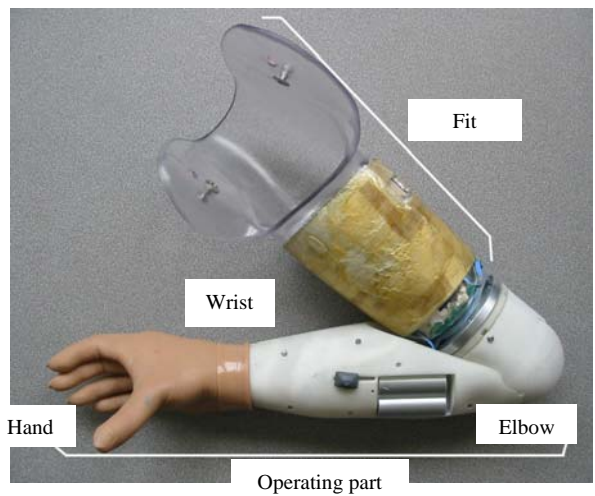

Fig. 10. The prosthesis of upper limbs developed

In this part, we have described the general structure of the prosthesis developed. The technologies used for the design and the manufacture have allowed to decrease the weight, to improve the aesthetic aspect and to reduce the production costs.

**III. The control card**

A general scheme of the control card is shown in figure 11. It represents the connections between the microcontroller and all the parts of the prosthesis. From this scheme, we can easily describe the functionality of the product developed.

The switch can activate or not the control card, which is alimented by a battery (7 V). The microcontroller can be connected to a computer, thanks to a serial link. It allows its programming.

At a first time, the two EMG electrodes measure a signal emitted by the contraction of the patient's healthy muscles. Secondly, this signal is treated and sent to the microcontroller (Figure 12). The treatment consists in recovering the envelope of the emitted signal.

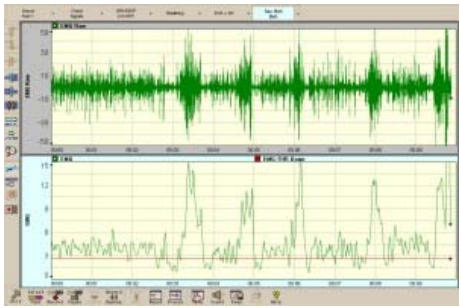

Fig. 12. EMG signal directly emitted by the patient's muscles and the treated signals

After control, the decision-making process allows the signal to be amplified, treated and sent to the corresponding motor. The choice of the motor is explain in the next part. There is one motor for each movement (elbow, wrist or hand) of the prosthesis.

Amplification and wave creation are used to inform the patient (thanks to an auricle) of the signal sent to the microcontroller. It is then easier for the patient to understand the use of its prosthesis. This function is currently under study [7].

**IV. The control of the prosthesis**

This prosthesis uses two electrodes developed and commercialized by OTTO BOCK. Currently, to use the OTTO BOCK or the UTAH prostheses, patients have to co-contract their muscles. This action is too difficult and needs training.

Currently, a simple adjustment of the control card makes possible to adapt the prosthesis to the sensitivity of the user and to regulate the delay with the emitted signals. This adjustment can be carried out simultaneously with the use of the prosthesis.

To product signals, the patient contracts the biceps or the triceps, which are the both healthy muscles it can use. With only two electrodes placed in the fit of the prosthesis, the patient can control the opening and the closing of the hand, the rotation of the wrist in the two directions and the extension or the inflection of the elbow. There is one motor for each movement.

Thanks to the decision-making process, the order was simplified; indeed no co-contraction is required to carry out a movement.

A strong contraction allows the motor selection. The movement is carried out by weak contractions. The selection is done as indicated hereafter:
- a strong contraction of the biceps implies the selection of the wrist motor (figure 13),
- a strong contraction of the triceps implies the selection of the elbow motor (figure 14).

After a delay, the control card selects automatically the opening/closing motor of the hand.

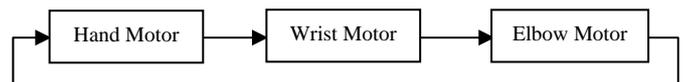

Fig. 13. Selection of the motor starting from a strong contraction of the biceps

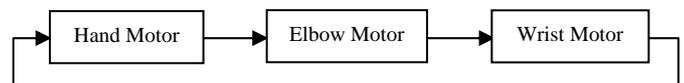

Fig. 14. Selection of the motor starting from a strong contraction of the triceps

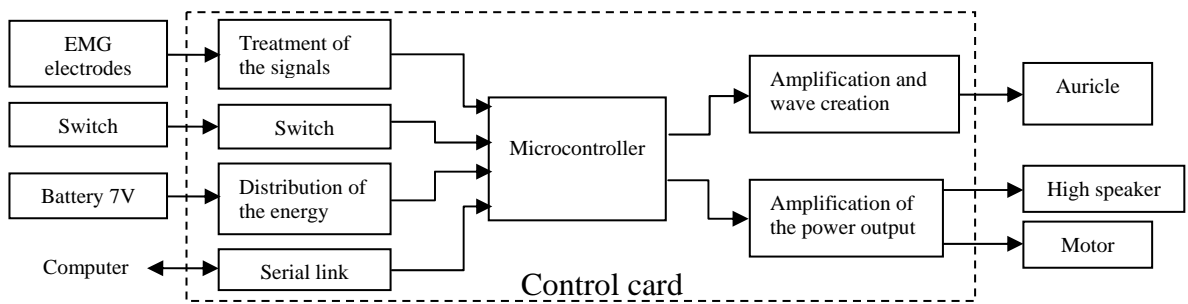

Fig. 11. General scheme of the control card

In the case of the hand motor is selected, a weak contraction of the biceps will open the hand and a weak contraction of the triceps will close it. The operations are the same for the others motors selected.

The figure 15 represents the diagram of the prosthesis motors activations according to the patient contractions of only two muscles (without co-contraction).

The identification of a strong or a weak contraction is explained in [8]. This study explains the treatment of the signal emitted by the muscles to be adapted to the decision-making process proposed. It deals with the main problems of recognition of the signal.

For a more effective and adapted use to the patients, we propose a protocol from which the top priority is to differentiate more quickly the strong signals from the weak signals and the dubious signals. We decided to work on the tangent at the origin of the sensors output signal (after treatment). Thus, by determining one time of contraction adapted to each patient, we can work on straight lines forms (figure 16).

We also introduce a strong slope limit and a weak slope limit (two reference slopes) in addition to a low limit and a high limit: if the starting slope of the signal is on the left of the strong slope limit, it is determined that it is a strong signal, whatever the end value taken by the top.

If it is on the right of the weak slope limit, it is determined that it is a weak signal, whatever the end value taken by the top.

Between the two slope limits, it is an indistinct signal. If the low limit is not exceeded, it is of a weak signal and if the high limit is exceeded, it is a strong signal.

With this decision-making process, the patient can control all its movements, only with two electrodes and without co-contractions. The prosthesis becomes more functional and easier to use.

Moreover, it is possible to evaluate and to detect the intention of movement of prosthesis [7]. The authors explain the sensibility of the signals emitted by the patient and its evolution with the time. They proposed different solutions to help the patient for the control of the prosthesis. The device combines an electric actuator associated with mechanical system, which allows the adaptability of the prosthesis.

### V. First Patient's return of use and opinion

This female patient used myoelectric prosthesis every day for more than 5 years. Her old prosthesis was constituted of a mechanical elbow and EMG control for wrist and hand. She uses the Tech.Innovation prosthesis for one month.

As she explains, the **aesthetics** is better now but it is awkward to let the prosthesis visible because of the aspect before the glove.

Compared to her old prosthesis, the **position of the battery** is satisfying (better than before), but the **mechanism of the manual positioning** of the elbow is less adapted. Moreover, the prosthesis is quiet; however the elbow is noisier than the hand.

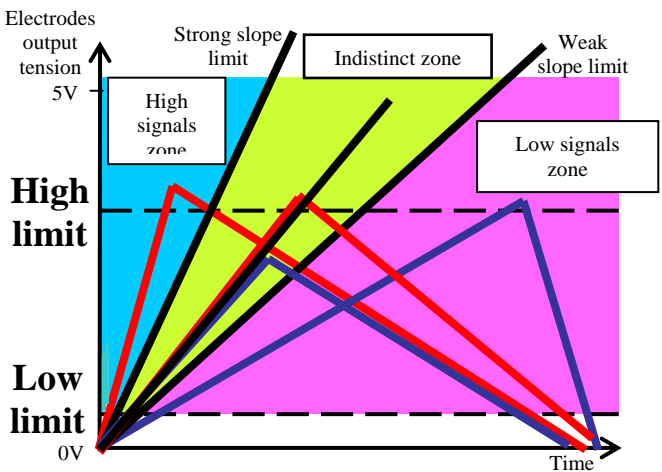

Fig. 16 Possible stimuli due to the contraction of a muscle and the way in which we can determine that they are strong stimuli or weak stimuli

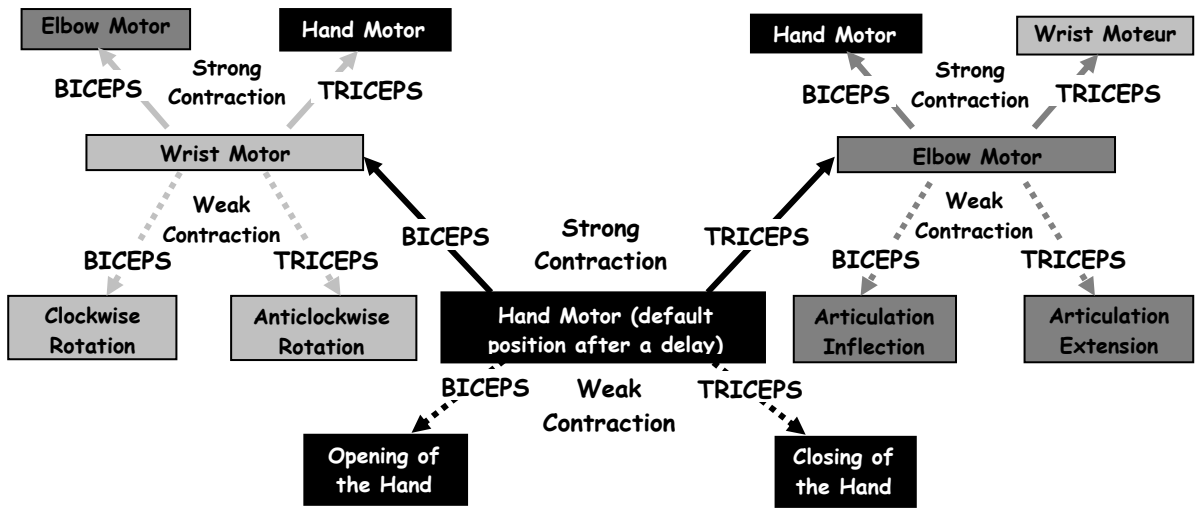

Fig. 15. Diagram of the prosthesis operations according to the muscular activity.

The patient is still learning to use this prosthesis, but after 15 days the practices were already present; she is very optimistic for the future. For the moment, she can not move the elbow as she wants (this is a new EMG control function) but the rotation of the wriest is easier than before: the preceding prosthesis used by the patient needed the triceps strong and weak contractions for the opening and closing of the hand and the biceps strong and weak contractions for the rotation of the wrist. The training of this new EMG elbow articulation control is apparently less easy than the "reuse" of the wrist and hand movements.

For the moment, there is no tiredness related to use of the prosthesis and no loss of the gestures precision at the end of the day.

The figures 17a and 17b show the patient equipped with the prosthesis developed by Tech.Innovation.

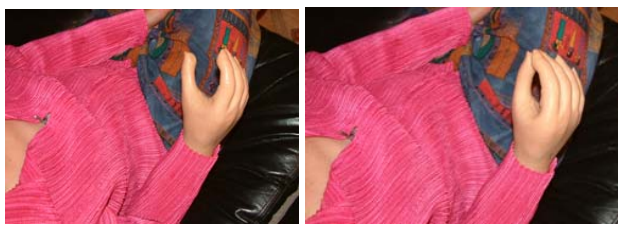

Figure 17a. Patient equipped with prosthesis: closing of the hand

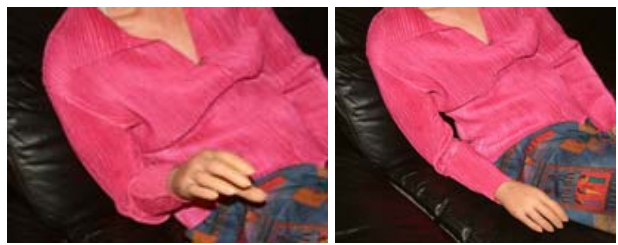

Figure 17b. Patient equipped with the prosthesis: extension of the elbow articulation

Even if only one patient currently uses the prosthesis and only since one month, we can conclude that the first remarks regarding the use of this new prosthesis are very encouraging.

## VI. Conclusion and Perspectives

In the medical world, a successful prosthesis is a prosthesis used. Today still 50% of the patients give up their prosthesis. A prosthesis must be functional, aesthetics, quiet, easy to use, more light and understood by the patient.

In this paper, we have exposed the different prostheses proposed by the main societies. We have described some of their defaults and some patient requirements.

With the technologies used to develop the prosthesis, we have tried to answer to the requirements of the amputees:
- the movements of the prosthesis allow a great diversities of actions. In the future, we can also add a kind of synthesized control, which allows the patient to make a complete action (several motors active at the same time),
- the aestheticism and the lightness of the prosthesis have been improved by the use of the stereolothography process and by the use of a commercial battery,
- the choice of the decision-making process simplifies the use of the prosthesis by the patient. They don't have co-contractions to do and the prosthesis reacts more quickly.
- a sonor feedback is installed [8] to allows the patient to correct its contraction if the prosthesis doesn't react as he wants.
- we develop a hand with five fingers. This product puts in action two additional fingers and allows the best catch (passive adaptability or passive compliance) : the fingers will come to marry the shape of the object.

Currently, we are designing a new control card. It integrates a fuzzy logic core, which allows a better piloting of the prosthesis. Throughout the day, the patient tires himself and thus, does not exert constant contractions. The fuzzy logic core allows the control card to adapt tiredness of the patient.

### Acknowlegment

We want to thank the Engineer School ESME SUDRIA, for the design of the control card and the integration of a Human-Machine Interface. We also collaborate with the centre of equipment of Gondreville and of Villiers Saint Denis. This collaboration enables us to collect the opinion of specialists (orthoprothesist) as well as the remarks of the patients concerning the prostheses.

Finally we also thank Professor Hamonet (Service de Médecine Physique et Réadaptation C.H.U. Henri MONDOR CRETEIL), Doctor Martinet and Doctor Deschamps (Responsibles of the Gondreville and Villier Saint Denis Orthopaedics Equipment Centres) for their assistances and their knowledge in the field of orthopedy.